\documentclass[aps,pre,array,epsfig,floatfix]{revtex4}
\usepackage{amsmath}
\usepackage{amsfonts}
\usepackage[]{subfigure}
\usepackage{color}
\usepackage{graphicx}
\usepackage{latexsym} 
\usepackage{amscd}
\usepackage{times}

\newcommand{\beq}{\begin{equation}}
\newcommand{\eeq}{\end{equation}}

\definecolor{mygray}{gray}{0.9}

\definecolor{mycyan}{cmyk}{0.0, 0.0, 0.0, 0.0}
\definecolor{mycolor}{RGB}{219, 0, 0}
\definecolor{yellow}{cmyk}{0.0, 0.0, 1.0, 0.0}

\newcommand{\texb}{\textcolor{blue}}

\newcommand{\fat}{\sffamily\textbf} 

\usepackage{soul}
\setstcolor{blue}

\begin{document}

\title{\fat{\texb{
Exclusion process on an open lattice with fluctuating boundaries.
}}}

\author{S. L. Narasimhan$^1$ and A. Baumgaertner$^{2,3}$}

\affiliation{
$^{1}$Chennai Mathematical Institute, SIPCOT IT Park, Siruseri - 603103, India\\and\\
$^{2,3}$Faculty of Physics, University of Duisburg-Essen, 47048 Duisburg, Germany}

\textit{•}

\begin{abstract}

We show that the TASEP of a driven system of particles of arbitrary size, with nearest neighbor repulsive interaction, on an open lattice is equivalent to the TASEP of interacting monomers on an open lattice whose size fluctuates in response to the entry and exit of particles. We have presented the maximal current profile as a function of the interaction strength for dimers and tetramers, obtained in Monte Carlo simulation; the results agree well with the ones computed by applying a specific rod-to-monomer mapping to the steady state current and density predicted by a mean-field theory of interacting monomers which adapts a Markov Chain approach for incorporating nearest-neighbor correlations. 
 
\end{abstract}

\date{\today}

\maketitle

\vspace{0.5cm}
\section{\bf Introduction}

A driven system of hard-core particles undergoing a unidirectional, stochastic motion in one dimension is a paradigm for studying the nature and evolution of non-equilibrium steady states \cite{Derrida1997,Derrida1998,Schutz2001b,Lipowsky2001,Klump2003,Klump2004,Chou2011,Dong2012}. Known as the totally asymmetric exclusion process (TASEP), the system is driven towards one of the three distinct non-equilibrium phases depending on the entry- and exit-rates, usually denoted by $\alpha$ and $\beta$ respectively; these phases are referred to as the Low-Density (LD), High-Density (HD) and the Maximal Current (MC) phases. The simplest model of such a system is provided by a one dimensional lattice-gas of hardcore particles with the following rules: (i) a given lattice-site can be occupied by at most one particle (hard-core exclusion); a particle that occupies only one lattice-site may be called a 'monomer';(ii) unidirectional hopping of a particle, say to the right, takes place only if the right nearest-neghbor site is vacant; (iii) a particle can enter (exit) the system only if the leftmost (rightmost) site is vacant (occupied).  Which stationary phase the system will be driven to is decided by the values of $\alpha$ and $\beta$. There is a point in the ($\alpha,\beta$)-phase diagram at which all the three phases meet - it may be called the \emph{triple-point}, $(\alpha ^*$,$\beta ^*)$, in analogy to what we may observe in an equilibrium system. Exact analysis of this model using a hydrodynamic approach \cite{Schutz1993,Schutz2000} on the one hand, and a matrix product method \cite{Derrida1992,Derrida1993} on the other, has firmly established the existence of the above mentioned three phases as long as the current-density relation is convex.\\

There have also been studies on the stationary states of a system of hard-core monomers with nearest-neighbor interaction, say $v$ in thermal units \cite{Pinkoviezky2013,Klump2004,Schutz2000,Hager2001}; in particular, Teimouri \emph{et al.,} \cite{Kolomeisky2015a} and D. C-Garza 
\emph{et al.,} \cite{Kolomeisky2015b} have proposed that the breaking and making of bonds between the nearest neighbors be treated as opposite chemical reactions implying a local detailed balance condition. Using Monte Carlo simulations and mean-field arguments, they have shown that such a system has a phase diagram consisting of the generic three distinct phases; the extent of these phases and the triple point depend on the nature (attractive or repulsive) and strength of the nearest neighbor interaction. More interestingly, the maximal monomer-current has been shown \cite{Kolomeisky2015a, Kolomeisky2015b} to have a peaking behaviour over a small range of repulsive interaction but approaches a value that corresponds to a system of non-interacting hard-rod dimers in the limit of very strong repulsion. This particular feature is indicative of correlations in the system.\\



The size of a particle in this model, $k$, is normally taken to be equal to the lattice constant (set equal to unity for convenience), which is also its hopping distance. It is quite possible to have a situation where the particle-size is different from its hopping distance. For example, protein synthesis in a living cell  are carried out by ribosomes which are large complex molecules that move along a messenger RNA (m-RNA)  reading  one codon (a triplet of three nucleotides) at a time for linking an amino acid to a growing chain of amino acids. Ribosomes are  complex molecules typically of size $\sim 20 nm$ that are large enough to cover approximately ten codons along m-RNA \citep{Heinrich1980, Kang1985}. In a lattice representation of m-RNA, each site of the lattice represents a codon; a ribosome on this lattice may be represented by a hard-rod particle covering many lattice sites but moving by one site at a time. In the discrete picture, therefore, $k$ need not be the same as its step-size or equivalently the lattice constant. Mean-field theoretical and Monte Carlo simulation studies \cite{Macdonald1968,MacDonald1969,Shaw2003,Lakatos2003} of unidirectionally driven systems of non-interacting ($v=0$) hard-rod $k$-mers have shown that these systems also have three distinct non-equilibrium phases; the location of the triple-point in the phase digram depends on the size, $k$, of the particle. Moreover, the density profile for $k>1$ in the HD phase has been shown \cite{Lakatos2003} to have an interesting non-monotonic branching feature at its exit end, unlike the one observed in the case of a monomer system.\\



Since these hard-rod particles may also be interacting with each other, it is of interest to study the stationary states of a driven system of interacting $k$-mers ($k\geq 1$) on an open one dimensional lattice-segment. Towards this end, discrete Takahashi lattice-gas model was proposed \cite{SLN2017} recently as a minimal mean-field description but it does not lead to the peaking of the maximal current at moderate strengths of repulsive nearest-neighbor interaction, which is what has been observed in the Monte Carlo studies \cite{Kolomeisky2015a, SLN2017}. Furthermore, it predicts a dependence of the triple-point on the interaction strength that is very different from what is observed in Monte Carlo simulation for extended particles ($k>1$). These discrepancies could be due to the fact that nearest-neighbor correlations and their influence on the local densities have been ignored in this mean-field description. \\ 

Since the density profiles in interacting particle systems will in general be inhomogeneous, it is necessary to relate the densities and the (nearest-neighbor) correlations on a local scale. Combined with a Markov Chain ansatz for the probability of a microstate that the system may assume, Dierl \emph{et al.,} \cite{Hager2001, Dierl2010, Dierl2012, Dierl2013} have shown that a time-dependent density functional theory, also referred to as the Markov Chain Adapted Kinetics (MCAK), can be used for studying the various boundary-induced phase transitions in interacting monomeric systems. This theory successfully reproduces the observed peaking behavior of the maximal current with respect to the nearest-neighbor (repulsive) interaction for monomeric systems. \footnote{An alternative 'cluster mean-field theoretical' (CMFT) approach to studying the stationary states of such interacting monomeric systems has recently been proposed by Kolomeisky and coworkers  \cite{Kolomeisky2017,Kolomeisky2018}, which is equivalent to the MCAK if only nearest-neighbor correlations are considered; see the Appendix of the archived article (S. L. Narasimhan and A. Baumbaertner, arXiv:1807.10667v1) for a discussion.}\\

Generalizing the MCAK theory for studying the stationary states of an interacting hard-rod $k$-mer system is however not a straightforward exercise; it will be simpler to see if a $k$-mer system can be mapped one-to-one into an interacting monomer system to which this theory applies. In this context, Gupta \emph{et al.,} \cite{Gupta2011} have shown that a system of non-interacting ($v=0$) hard-rod $k$-mers undergoing TASEP on a lattice-ring has a homogeneous stationary state and hence can be mapped uniquely into a system of monomers undergoing exactly the same dynamics on a lattice-ring of reduced size. Since the number of $k$-mers in the system and hence their representative monomers is constant, the (reduced) size of the representative lattice-ring is also a constant. Hence, the mapping they have proposed is exact on a lattice-ring. \\

In the case of an open system on the other hand, the size of the representative lattice will fluctuate in response to the random entry and exit of $k$-mers; this is so because the first $k$ sites occupied by the newly entered $k$-mer will be represented by a single occupied site implying thereby a decrease in the lattice-size by $(k-1)$; similarly, the exit of a $k$-mer, if represented by a single occupied site, will increase the system-size by $(k-1)$. The Monte Carlo simulation of this model describing an exclusion process on an open lattice with fluctuating boundaries (EPFB) demonstrates that the lattice-size evolves towards a stationary value in the same way as the number of representative monomers in the system does. Our simulation clearly shows that the rod-to-monomer mapping of Gupta \emph{et al.,} \cite{Gupta2011} holds good if the steady-state values of the lattice-size and the number of monomers are used in stead of constant values as on a lattice-ring. More interestingly, we observe that the steady-state behavior of a $k$-mer system in this EPFB representation is the same as what we would compute from the MCAK model by using the exact rod-to-monomer mapping of Gupta \emph{et al.,} \cite{Gupta2011}. We describe this size-fluctuating EPFB model in the next section and present our simulation data in section III; finally we discuss and summarize the results in the last section. \\



\section{\bf The Model.}

Let there be $M$ $k$-mers on a one dimensional lattice-segment, say $\cal{L}_N$, that consists of $N$ sites ($N \geq Mk$); by a $k$-mer, we mean a particle that occupies $k$ sites of the lattice. Let the position of a particle, say $x$, be taken to be the lattice-site occupied by its right end. If $x_l$ is the position of the $l^{th}$ particle, we have the ordered configuration, $\{ x\}_M = \{x_1 < x_2 < \cdots < x_M\}$ such that the distance between the nearest neighbours is never less than $k$, \emph{i.e.,} $r_l \equiv x_l - x_{l-1} \geq k$ for $l = 2,3,\cdots ,M$. We define the distances, $r_1 \equiv x_1 \geq k$ and $r_{M+1} \equiv N-x_M$ so that we have the constraint $\sum _{l=1}^{M+1} r_l = N$. At any instant of time, the system is completely specified by the ordered set of coordinates, $\{x\}_M$, also referred to as its \emph{microstate}.\\

The stochastic evolution of the system in the bulk (\emph{ignoring boundary effects}) is described by the Master Equation,
\begin{equation}
\frac{\partial P(\{x\}_M;t)}{\partial t} = \sum _{\{x\}'_M} \left[ \omega ({\{x\}'_M}\rightarrow {\{x\}_M})P(\{x\}'_M;t) - \omega ({\{x\}_M}\rightarrow {\{x\}'_M})P(\{x\}_M;t)\right]
\label{Eq:MEq1}
\end{equation}
where $P(\{x\}_M;t)$ is the (normalized) probability that the system is in the microstate $\{x\}_M$ at time $t$; $\omega$ is the transition-rate from one microstate to another, which specifies the rate at which a randomly chosen particle hops to its vacant right nearest neighbor site.\\

In the presence of nearest-neighbor interactions, the hopping rates of a $k$-mer will depend on where its nearest neighbours are located on the lattice. Let $v(r_l)$ be the interaction between the $l^{th}$ and $(l-1)^{th}$ particles, expressed in $kT$ units. We also set $v(r_1)=0=v(r_{M+1})$ because the first (last) particle does not have a left (right) nearest neighbor. We have a system of sticky rods described by the Takahashi Hamiltonian,
\begin{equation}
H = \sum_{l=1}^{M+1} v(r_l)
\label{Eq:Takahashi}
\end{equation}
Treating the  particles as sticky rods of length $k$, we have the following specific form for $v(r)$:
\begin{equation}
v(r) = \left\lbrace \begin{array}{l l} \infty & \quad \text{if $r<k$}\\
                                       v & \quad \text{if $r=k$}\\
                                       0 & \quad \text{if $r>k$}\\
              \end{array}
       \right.
\label{Eq:sticky}
\end{equation}
As illustrated in Fig.(\ref{Fig:Model}) for trimers, the hopping of the $l^{th}$ $k$-mer from position $x_l$ to $x_l+1$ depends on the the positions of its nearest neighbors ($x_{l-1} \leq x_l-k$ and $x_{l+1} \geq x_l+k+2$).\\

Therefore, assuming local detailed balance condition as in references \cite{Kolomeisky2015a,Kolomeisky2015b}, the hopping rate of say the $l^{th}$ $k$-mer can be witten as
\begin{equation}
\omega _l(x_{l-1},x_{l+1}) = \left[ 1+\delta _{x_{l-1},x_l-k}(e^v-1) \right]
                             \left[ 1-\delta _{x_{l+1},x_l+k+1} \right]
                             \left[ 1+\delta _{x_{l+1},x_l+k+2}(e^{-v}-1)\right]
\label{Eq:HopRate}
\end{equation}
where the middle term ensures that the site $x_l+1$ is unoccupied. 
The transition-rate $\omega$ from one microstate to another is then given by the sum,
\begin{equation}
\omega ({\{x\}_M}\rightarrow {\{x\}'_M}) = 
                \sum _l \delta _{\{x\}'_M, \{x\}^l_M}\omega _l(x_{l-1},x_{l+1})
\label{TRate}
\end{equation}
where $\{x\}^l_M$ is a configuration in which the $l^{th}$ particle is at site $x_l+1$. This, along with the Master Equation, Eq.(\ref{Eq:MEq1}), suggests that the average stationary current due to the hopping of the $l^{th}$ particle to its right nearest neighbor site, $x_l+1$, is given by
\begin{equation}
j_l = \langle \left[ 1+\delta _{x_{l-1},x_l-k}(e^v-1) \right]
                             \left[ 1-\delta _{x_{l+1},x_l+k+1} \right]
                             \left[ 1+\delta _{x_{l+1},x_l+k+2}(e^{-v}-1)\right]
      \rangle
\label{Eq:Currentl}                             
\end{equation}
where $\langle \cdots \rangle$ denotes averaging with respect to the steady state probability distribution, $P(\{x\}_M)$. Clearly, $j_l$ has contributions from the nearest-neighbor correlations. This becomes more transparent if we describe a microstate of the system in terms of the occupancies of the lattice sites rather than in terms of the positions of particles. A simplified but dynamics-preserving version of such a description requires that a $k$-mer be treated as a single particle occupying a single site of a lattice rather than as a row of $k$ consecutive occupied sites, as described in \cite{Gupta2008, Gupta2011}.\\

\subsection{Mapping of a $k$-mer system into a monomer system.} 

Let there be $M$ $k$-mers located at positions, $\{x\}_M$, on an open lattice-segment, say $\cal{L}_{N}$, of $N$ sites. The number of empty sites on this lattice is therefore $(N-Mk)$. Let us now consider a lattice segment, $\tilde{\cal{L}}_{\tilde{N}(M)}$, which consists of $\tilde{N}(M)\equiv (N-Mk+M)$ sites of which $M$ sites are occupied in the same order as the $k$-mer positions, $\{x\}_M$, on $\cal{L}_N$; the empty sites of $\tilde{\cal{L}}_{\tilde{N}(M)}$ correspond to those of $\cal{L}_{N}$ again in the same order. Thus, every $k$-mer on $\cal{L}_N$ is represented by a single occupied site (a monomer) in $\tilde{\cal{L}}_{\tilde{N}(M)}$, and every empty site of $\cal{L}_N$ corresponds to an empty site of $\tilde{\cal{L}}_{\tilde{N}(M)}$.\\

\begin{figure}  
\begin{center}

\vspace{-2cm}
\includegraphics[scale=0.9]{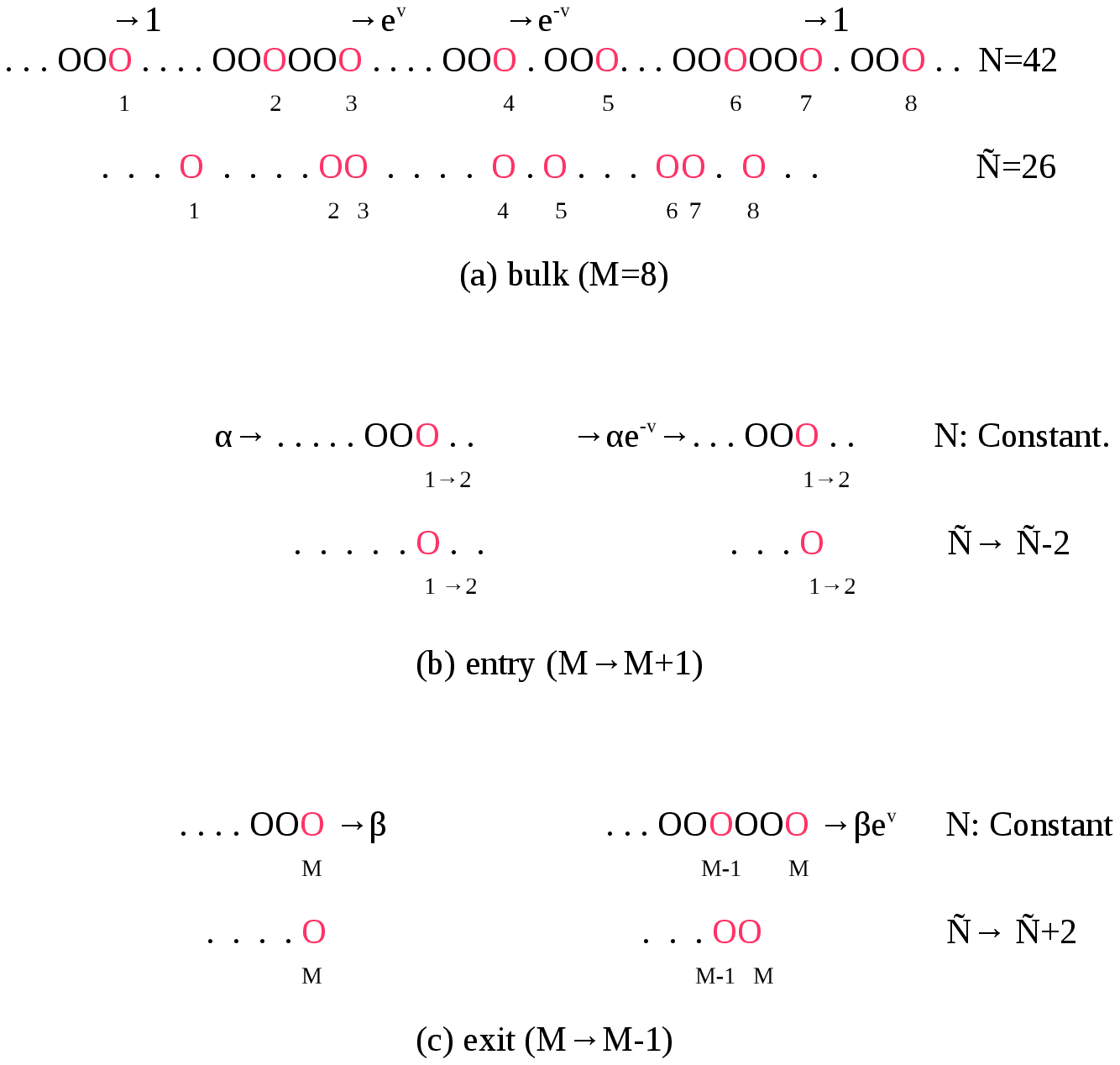}
\vspace{-9cm}

\caption{Mapping of a trimer-syetem onto a monomer system in the BFEP model. The position of a trimer is identified with the site occupied by its right edge, represented by a red circle. (a) Bulk: The number of sites in the \emph{reduced} lattice is $\tilde{N}=26$ which is equal to the sum of the empty sites ($=18$) and the number of trimers ($M=8$) in the original lattice  of $N=43$ sites. The transition rates for various situations, marked on top, are the same for both the lattices. (b) Entry: On entry of a new trimer, at the rate $\alpha$ or $\alpha e^{-v}$ as the case may be, the old one labelled 1 is relabelled 2. The size of the \emph{reduced} lattice is decreased by 2. (c) Exit: When the last particle, labelled $M$ exits, the size of the \emph{reduced} lattice is increased by 2.}
\label{Fig:Model}

\end{center}
\end{figure}

Fig.(\ref{Fig:Model}) schematically illustrates this mapping of a trimer system to a monomer system. There is one-to-one correspondence between the empty sites of $\cal{L}_N$ and those of $\tilde{\cal{L}}_{\tilde{N}(M)}$ in the bulk but not in the entry- and exit-regions. For example, as illustrated in the subfigure (b) of Fig.(\ref{Fig:Model}), a trimer entering the system will occupy the first three empty sites in the entry-region of $\cal{L}_N$, while its representative monomer in $\tilde{\cal{L}}_{\tilde{N}(M)}$ occupies only one site; however, the mapping preserves the number of empty sites between particles. If the leftmost particle is always labelled 1, then $\tilde{N}(M)\to [\tilde{N}(M) -2]$. Similarly, as shown in the subfigure (c), the exit of a trimer from $\cal{L}_N$ corresponds to the exit of a monomer from $\tilde{\cal{L}}_{\tilde{N}(M)}$ resulting in an increase, $\tilde{N}(M)\to [\tilde{N}(M) +2]$. In other words, the size $\tilde{N}(M)$ of the \emph{reduced} lattice will fluctuate - \emph{i.e}., $\tilde{N}(M)\to [\tilde{N}(M) \mp (k-1)]$ - in response to the random entry and exit of $k$-mers. This process may be called an \emph{Exclusion Process on an open lattice with Fluctuating Boundaries} (EPFB) so as to distinguish it from other types of exclusion processes.\\



At any point of time during the evolution of the system, the number of particles, $M(t)$, is the same on both $\cal{L}_{N}$ and $\tilde{\cal{L}}_{\tilde{N}(M)}$; the corresponding densities are $M(t)/N$ and $M(t)/\tilde{N}(M)$ respectively. Since $M(t)$ is a stochastic variable, we have the configuration-averaged densities $\rho_k(t)\equiv \langle M(t)\rangle /N$ and $\tilde{\rho}(t) \equiv \langle M(t)/\tilde{N}(M)\rangle$ respectively. On the basis of the Monte Carlo evidence that $\langle M(t)/\tilde{N}(M)\rangle$ is equal to $\langle M(t)\rangle /\langle \tilde{M}(N)\rangle$ within statistics - presented in the next section -  we have
\begin{equation}
\tilde{\rho}(t) = \frac{\langle M(t)\rangle}{N-\langle M(t)\rangle (k-1)}\equiv 
                  \frac{\rho_k(t)}{1-\rho_k(t)(k-1)}
\end{equation}
or equivalently,
\begin{equation}
\rho_k(t) = \frac{\tilde{\rho}(t)}{1+\tilde{\rho}(t)(k-1)}
\label{Eq:k-Dens}
\end{equation}
which is exactly the mapping proposed by \cite{Gupta2008, Gupta2011} for particles on a lattice-ring. The difference is that $\tilde{\rho}(t)$ is the average density of representative monomers on a \emph{reduced} lattice, $\tilde{\cal{L}}_{\tilde{N}(M)}$, whose size fluctuates in response to the entry and exit of particles. Moreover, by construction, the hopping dynamics on $\tilde{\cal{L}}_{\tilde{N}(M)}$ is the same as on $\cal{L}_{N}$ and therefore we expect the $k$-mer current, $j_k$, on $\cal{L}_{N}$ to be related to the monomer-current, $\tilde{j}$, on $\tilde{\cal{L}}_{\tilde{N}(M)}$ by the same mapping,
\begin{equation}
j_k = \frac{\tilde{j}}{1+\tilde{\rho}(t)(k-1)}
\label{Eq:k-Curr}
\end{equation} 

If we now denote the occupancy of a site in $\tilde{\cal{L}}_{\tilde{N}(M)}$ by $n = 0$ or $1$, then the hopping rate for a monomer occupying the site $l$ may be written as
\begin{equation}
\omega _l(n_{l-1},n_{l+2}) = n_l (1-n_{l+1})\exp [-v(n_{l+2}-n_{l-1}] \\
\end{equation}
which may be rewritten in the equivalent form,
\begin{equation}
\omega _l(n_{l-1},n_{l+2}) = n_l (1-n_{l+1})[1+n_{i-1}(e^v-1)][1+n_{l+2}(e^{-v}-1)]
\label{Eq:HopRateMono}                  
\end{equation}
Hence, the steady-state current, $\tilde{j}_{l\to l+1}$ through the bond, $(l,l+1)$:
\begin{equation}
\tilde{j}_{l\to l+1} = \langle \omega _l(n_{l-1},n_{l+2})\rangle
\end{equation}
where the averaging is done with respect to the steady state probability distribution, $P(n_1,n_2,\cdots,n_{N(M)})$. When expanded, this leads to the following expression,
\begin{equation}
\hat{j}_{l\to l+1} = 
              \langle \tilde{n}_{l-1}n_l\tilde{n}_{l+1}\tilde{n}_{l+2}\rangle +
              \langle \tilde{n}_{l-1}n_l\tilde{n}_{l+1}n_{l+2}\rangle e^{-v} +
              \langle n_{l-1}n_l\tilde{n}_{l+1}\tilde{n}_{l+2}\rangle e^v +
              \langle n_{l-1}n_l\tilde{n}_{l+1}n_{l+2}\rangle
\label{Eq:CurrentlMono}              
\end{equation}
where $\tilde{n} \equiv (1-n)$. Since we are dealing with a non-equilibrium process, we need an exact functional form for the probability $P(n_1,n_2,\cdots,n_{N(M)})$ ( $ = P(\{n\})$, for notational convenience) in order to compute the various averages. Dierl \emph{et al}., have proposed a procedure, called Markov Chain Adapted Kinetics (MCAK) \cite{Dierl2013},for computing these averages for an interacting system of monomers. We use their expression for the steady-state monomer current, along with Eq.(\ref{Eq:k-Curr}), to see how it compares with the current obtained by from BFEP simulation.\\

\section{Monte Carlo results }.

We have used the continuous-time algorithm of Bortz, Kalos and Lebowitz \cite{Bortz1975} to simulate the steady state behavior of an interacting system of $k$-mers and its monomer representatives. This algorithm, also known as the Kinetic Monte Carlo (KMC) algorithm, has been used for studying reaction-diffusion processes as well.\cite{Gillespie1976, Gillespie2001}. \\

(a) Consider an arbitrary configuration of $k$-mers, say $\{x_j | j=1,2,\cdots M\}$ where $x_j$ is the position of the left-end of the $j^{th}$ $k$-mer. If the first $k$ sites are empty, we set the entry-rate $w_0 = \alpha $ or $q\alpha$ depending on whether the site $k+1$ is empty or occupied; here, $q\equiv e^{-v}$. Subsequently, every $k$-mer in the interior of the system that can jump to its empty right nearest-neighbor site can be assigned one of the rates, $ w_j = 1,q$ or $r (\equiv 1/q)$ depending on whether it makes or breaks a nearest-neighbor contact; the exit-rate for a $k$-mer is either $\beta$ or $r\beta$ depending on whether it is isolated or not; if it is not  isolated, it exits the system by breaking a nearest-neighbor bond.\\ 

(b) However, in the EPFB model, occupying the leftmost site of $\tilde{\cal{L}}_{\tilde{N}(M)}$ represents the entry of a $k$-mer and so, it is done if the first $k$ sites are empty; the entry-rate is $\alpha $ or $q\alpha$ depending on whether the $(k+1)$-th site is empty or occupied. Simultaneously, the other monomers in the system are renumbered $j = 2, 3, \cdots M+1$ and their positions, $x_j$ are reset to $[x_j - (k-1)]$. This amounts to decreasing the size of the lattice by $(k-1)$. On the other hand, the exit of a monomer represents the exit of a $k$-mer; so, the size of the lattice is increased by $(k-1)$ and there is no need either to renumber the monomers or to reset their positions. Again, the exit-rate is $\beta$ or $r\beta$ depending on whether the site $[N(M)-1]$ is empty or occupied. A monomer hopping to its empty right nearest-neighbor site in the interior of the lattice is assigned a rate $0,1,q$ or $r$ depending on whether it is isolated, or it makes or breaks a nearest-neighbor bond. This procedure is illustrated in Fig.(\ref{Fig:Model}) for a trimer system. \\

In either case, we compute the cumulative rates, $R_j \equiv \sum _{i=0}^j w_i$, for $j = 0, 1,2, \cdots M$. Which particle is going to make a jump, and how much time may have elapsed for this event to occur are decided by the following strategy. Let $n_1$ and $n_2$ be two uniform random numbers in the range $(0,1)$. If $n_1 R_M \leq R_0$, then a particle enters the system; else, the value of $l$ for which the inequality $R_{l-1} < n_1 R_M \leq R_l$ holds points to the particle that makes a jump. Under the implicit assumption that it is a Poisson process, the time associated is given by $-\ln (n_2)/R_M$. This way of choosing a particle to make a jump is repeated a large number of times, say $\sim 10^7-10^8$ times, in a single run. The current-profile is obtained as the ratio of the number of times a particle entered (or equivalently, exited) a particular site, $j$, to the cumulative total of the times associated with these events. This may be averaged over a number of runs as well. Our focus is mainly on the \emph{reduced} system which is a representation of an open $k$-mer system.\\

\subsection{Non-interacting ($v=0$) particles.}

In the absence of nearest-neighbor interaction ($v=0$), the EPFB model describes a monomer system whose steady-state behavior is controlled by two parameters, $\alpha$ and $\beta$. In the simulation, however, we set $\alpha+\beta=1$ so that we have a single tuning parameter. We start with an empty lattice of size $L_0=500$; as the system evolves, the number of particles and the corresponding lattice sze will change. We find that $10^7$ dynamic events (entry, exit and hopping)are enough to leave the system in the steady-state; this constitutes one single run. The Monte Carlo data presented here are an average of 50 independent runs.\\

\begin{figure}  
\begin{center}

 \includegraphics[width=0.7\textwidth]{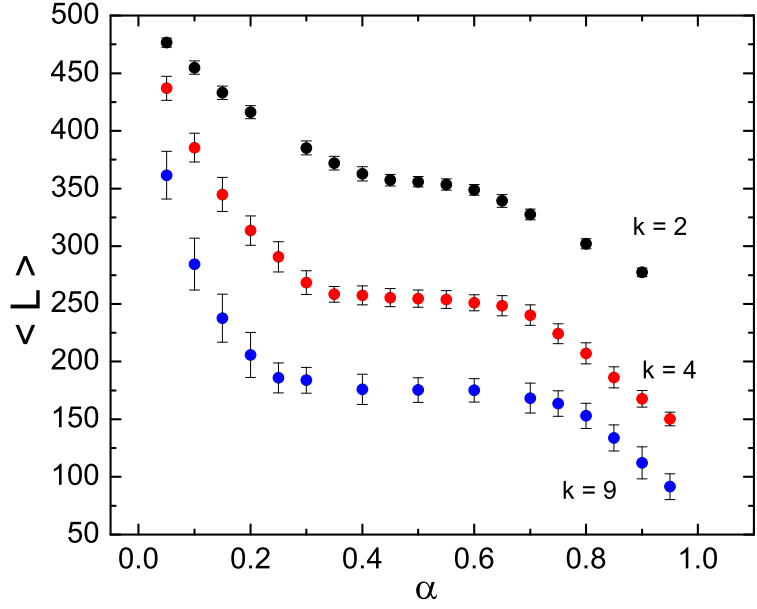}
 
  \caption{Steady-state system size as a function of $\alpha$ for $k=2, 4$ and $9$ in the BFEP model, illustrated in Fig.(\ref{Fig:Model}). Initial system size $L_0=500$ and the initial number of particles in the system $N_0=0$. In the simulation, $\alpha+\beta=1$. In the steady-state, we have the equality, $L(N)+N(k-1)=L_0$. Data are an average of 50 runs each consisting of $10^7$ hopping events. } 
  \label{Fig:SystSize}

\end{center}
\end{figure}

In Fig.(\ref{Fig:SystSize}), we have presented the steady-state values of the system size, $L(N)$, for $k=2, 4, 9$ as a function of $\alpha$ in the range $(0,1)$. The plateau region corresponds to the Maximum Current (MC) phase. We find that the equality, $L(N) = L_0 - N(k-1)$, holds good whatever be the value of $\alpha$. In fact, even if the lattice were not empty initially, by construction, the initial size of the lattice will be $L(N_0) = L_0 - N_0(k-1)$ and the equality has been found to hold good in the staedy-state. We see that the error on the data points increase with the rod-size, $k$. For example, it is almost the size of the data point for $k=2$ while it is comparatively large in the case of $k=9$ - data for small values of $\alpha$ carry larger error than data for large values of $\alpha$; at $\alpha = 0.05$, it is roughly $12\%$ for $k=9$.\\

\begin{figure} 
\begin{center}

 \includegraphics[width=0.7\textwidth]{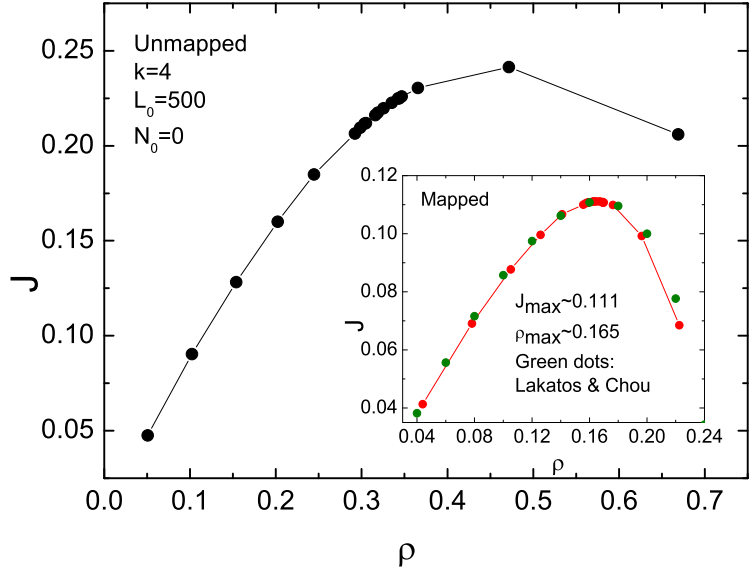}
  \caption{Black dots represent the Current-Density ($\tilde{J}-\tilde{\rho}$) data  for $k=4$ in the BFEP model. The corresponding data for tetramers, obtained by using Eq.(\ref{Eq:k-Dens}) and Eq.(\ref{Eq:k-Curr}) are shown in the inset. The green dots in the inset are obtained by using the Lacatos-Chou formula, $J_4=(\rho(1-4\rho)/(1-3\rho)$. The agreement is quite good, and within ststistics.} 
\label{Fig:Curr-Dens-k4}   

\end{center} 
\end{figure}

The steady-state monomer density, $\tilde{\rho} = N/L(N)$, corresponds to the $k$-mer density, $\rho_k$, given by Eq.(\ref{Eq:k-Dens}); similarly, the steady-state monomer current, $\tilde{J}$, corresponds to the $k$-mer current, $J_k$, given by Eq.(\ref{Eq:k-Curr}). In Fig.(\ref{Fig:Curr-Dens-k4}), we have presented (black dots) the current-density relation for $k=4$. In the inset of this figure, we have presented the corresponding $k$-mer current-density relation. It is asymmetric and its peak value, $J_4 \sim 0.11$, corresponds to the density, $\rho_4 \sim 0.165$. These values agree very well with the expected values $1/9$ and $1/6$ \cite{Lakatos2003} respectively. In fact, the green dots in the inset correspond to the Lacatos-Chou relation, $J_4 = \rho_4(1-4\rho_4)/(1-3\rho_4)$ \cite{Lakatos2003}.
The agreement with the simulation data is quite good. \\

In Fig.(\ref{Fig:Curr-alfa-k249}), we have presented the $k$-mer current obtained using Eq.(\ref{Eq:k-Curr}) as a function of $\alpha$ for $k=2, 4, 9$. The plateau regions correspond to the MC-phase, which sets in at $\alpha \sim 0.4, 0.32, 0.23$ for $k=2, 4, 9$ respectively. The larger values of $\alpha$ at which $J$ starts dipping correspond to $(1-\beta) \sim 0.6, 0.68, 0.77$ and so we have the equality $\alpha=\beta$. This is, by definition, the \emph{triple-point} where all the three phases meet. The above estimated values of the \emph{triple-point} agree very well with the theoretically expected values for $k$-mers, $\alpha=\beta=1/(\sqrt{k}+1)$ \cite{Lakatos2003}.\\

We may therefore conclude that the EPFB model is a good representation of a system of hard-rod $k$-mers undergoing a TASEP on an open lattice.

\subsection{Interacting ($v>0$) particles.}

An interesting feaure of the current profile for a $k$-mer system with respect to the strength of nearest-neighbor repulsion ($v>0$) is the peaking of the current at small values of $0 < v \lesssim 1$ \cite{Kolomeisky2015a, SLN2017}. In the limit of very strong repulsion $v \gg 1$, the current has the same value as that for a hard-rod $(k+1)$-mer. How the current profile of the EPFB model compares with that of a standard TASEP model is of interest.\\

\begin{figure}  
\begin{center}

 \includegraphics[width=0.7\textwidth]{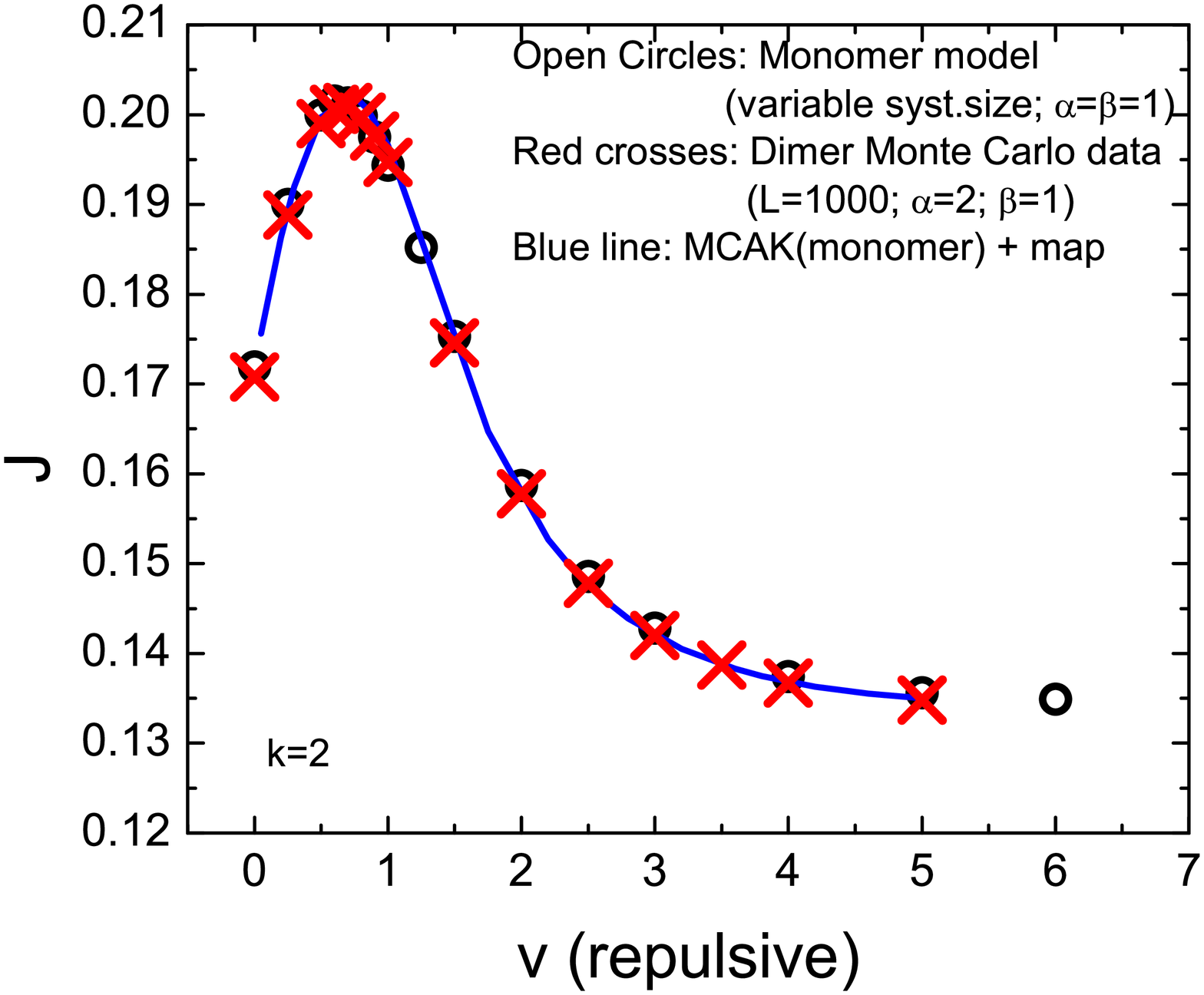}
  \caption{Current profile in the MC-phase for $k=2$: The open circles are the dimer current estimtes obtained from the monomer current of the BFEP model, with $\alpha=\beta=1$, by using Eq.(\ref{Eq:k-Curr}. The red crosses are the Monte Carlo estimates of the dimer current in a standard TASEP model on a lattice of fixed size, $\cal{L}_{N}$ (in the text); $L=1000, \alpha=2, \beta=1$. The continuous blue line is dimer current obtained from the MCAK monomer current by using Eq.(\ref{Eq:k-Curr} where the density is for a lattice of fixed size. } 
\label{Fig:Curr-Dimer-v}   

\end{center} 
\end{figure}

\begin{figure}  
\begin{center}

 \includegraphics[width=0.7\textwidth]{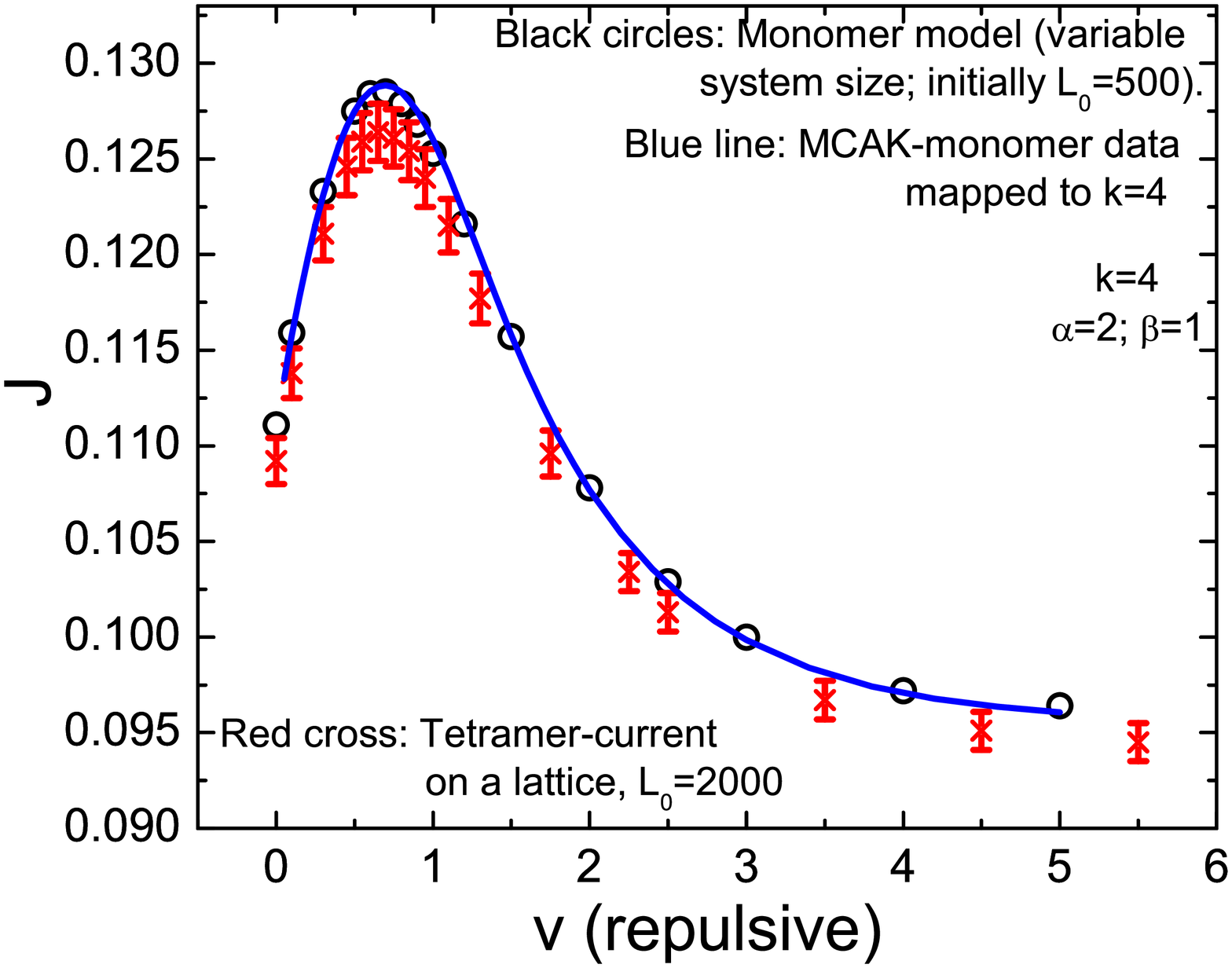}
  \caption{Current profile in the MC-phase for $k=4$: The open circles are the tetramer current estimtes obtained from the monomer current of the BFEP model, with $\alpha=2, \beta=1$, by using Eq.(\ref{Eq:k-Curr}. The red crosses are the Monte Carlo estimates of the tetramer current in a standard TASEP model on a lattice of fixed size, $\cal{L}_{N}$ (in the text); $L=2000, \alpha=2, \beta=1$. The continuous blue line is tetramer current obtained from the MCAK monomer current by using Eq.(\ref{Eq:k-Curr} where the density is for a lattice of fixed size.}   

\label{Fig:Curr-kp4-v} 
\end{center} 
\end{figure}

In Fig.(\ref{Fig:Curr-Dimer-v}), we have presented as open cicles the dimer current profile obtained from the EPFB monomer current by using Eq.(\ref{Eq:k-Curr}). The parameters used in the EPFB simulation are $L_0=500; \alpha=\beta=1$. Note that the system size $L(N)$ in EPFB is a dynamically fluctuating variable in response to the entry and exit of particles. In the same figure, we have also presented (red crosses) the Monte Carlo estimates of the dimer current in a standard TASEP simulation with parameters $L=1000, \alpha=2, \beta =1$. Here, $L$ is fixed. There is a good agreement between these two profiles  over the range of $v$ studied. Similarly, for tetramers ($k=4$) shown in Fig.(\ref{Fig:Curr-kp4-v}), the agreement is reasonably good even though the EPFB seems to slightly over-estimate the current values as compared to those of the standard TASEP; the maximum difference in their values in the peak-region is $\sim 0.003$. There is good agreement between these two cases in the extreme limits, $v=0, \gg 1$.\\

In these figures, the continous blue line is the current profile generated by mapping the steady-state MCAK current given below into that of a tetramer by using Eq.(\ref{Eq:k-Curr}):
\begin{equation}
j(\rho)  =  \frac{[\rho-C(\rho))]^2}{\rho(1-\rho)}
            \left[1-2[\rho-C(\rho)](1-e^{-v})
            \right]
\label{Eq:MCAK-Current}
\end{equation}
where $C(\rho)$ is the nearest-neighbor correlation given by
\begin{equation}
C(\rho) = \frac{e^{-2v}[\rho-C(\rho)]^2}{1-2\rho+C(\rho)}
\label{Eq:MCAK-Corr}
\end{equation}
What value of $\rho$ has to be used in order to compute the maximal current needs a careful consideration because the current-density relation has two maxima for $v > v_c \sim 1.44$ separated by a minimum at $\rho=1/2$. The densities, $\rho_{1,2}$ corresponding to these two maxima depend on the value of $v$. In particular, in the limit $v\to \infty$, $\rho_{1,2} = (1-1/\sqrt{2},1/\sqrt{2}) \equiv (1/[\sqrt{2}(\sqrt{2}+1)],1/\sqrt{2})$. It may be noted that $\rho_1$ is the density at which the current for a hard dimer system has a maximum value, as is immediately clear from the Lakatos-Chou expression for the hard-rod $k$-mer current \cite{Lakatos2003},
\begin{equation}
j_k = \frac{\rho_k(1-k\rho_k)}{[1-(k-1)\rho_k]}
\label{Eq:LC-Curent}
\end{equation}
which has a maximum value $j_k = [\sqrt{k}+1]^{-2}$ at $\rho_k = [\sqrt{k}(\sqrt{k}+1)]^{-1}$. This suggests that we must choose the value for $\rho$ in Eqs.(\ref{Eq:MCAK-Current},\ref{Eq:MCAK-Corr}) that corresponds to the first maximum of the MCAK current for a given value of $v$. We then use Eq.(\ref{Eq:k-Curr}) to compute the corresponding tetramer current. The MCAK current profiles thus computed, shown as continuous blue lines in Figs.(\ref{Fig:Curr-Dimer-v},\ref{Fig:Curr-kp4-v}) agree very well with those (open circles) obtained from EPFB simulation.\\

\section{Discussion.}

\begin{figure}  
\begin{center}

 \includegraphics[width=0.7\textwidth]{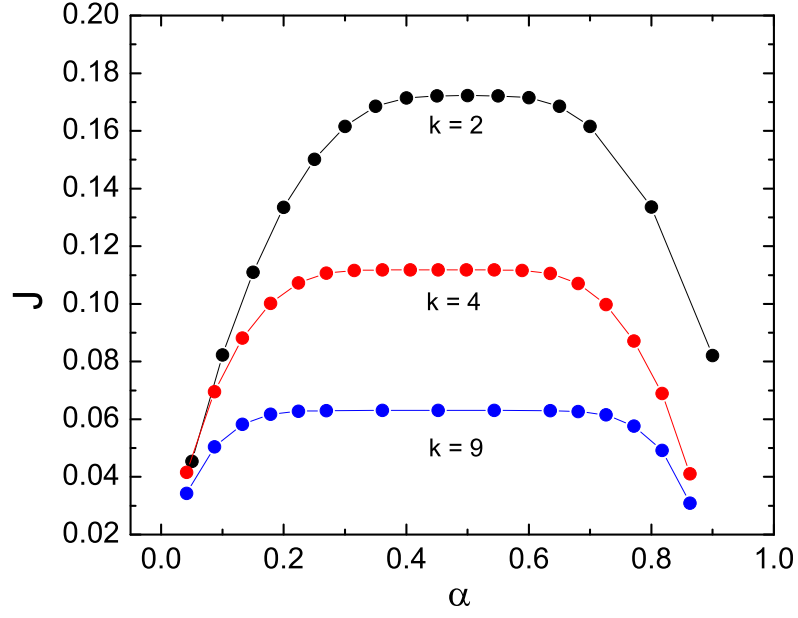}
  \caption{The current, $J$, as a function of $\alpha$ for $k=2, 4$ and $9$, obtained by using Eq.(\ref{Eq:k-Dens}) and Eq.(\ref{Eq:k-Curr}). The plateau regions correspond to the Maximal Curent (MC) phase, which sets in at values $\alpha=\beta \sim 0.4, 0.32, 0.23$ for $k=2, 4, 9$ respectively. These are in very close agreement with the expected values, $1/[\sqrt{k}+1]$ \cite{Lacatos2003} }. 
\label{Fig:Curr-alfa-k249}   

\end{center} 
\end{figure}

In the case of non-interacting ($v=0$) particles, the steady-state current and density values obtained from EPFB simulation, when substituted in Eq.(\ref{Eq:k-Curr},\ref{Eq:k-Dens}), agree quite well with the mean-field predictions of Lakatos and Chou for $k$-mers \cite{Lakatos2003}. Even the triple-point estimates, (see Fig.(\ref{Fig:Curr-alfa-k249}), are fairly accurate. This suggests that the TASEP of a hard-rod $k$-mer system is equivalent to the TASEP of a monomer system whose (lattice) size fluctuates in response to the entry and exit of particles. The steady-state values of the system-size, $L(N)$, and of the number of monomers in the system, $N$, satisfy the equality $L(N)+(k-1)N=L_0$ where $L_0$ is the initial size of the lattice. If we want to start the process with $N_0$ monomers in the system, then the initial system-size is set equal to $L_0-(k-1)N_0$ in the EPFB algorithm. Clearly, $N \geq L_0/(k-1)$, at any stage during the evolution of the system. \\

The maximal current profile with respect to the strength of repulsive interaction, $v \geq 0$, generated by the EPFB model agrees reasonably well with the one generated by the standard TASEP simulation of $k$-mers, especially for small values of $k$. However, we do observe that, for $k=4$, the EPFB profile is slightly overestimated as compared to that obtained by the standard TASEP simulation - the maximum difference is in the peak region and is $\sim 0.003$. On the other hand, the EPFB current profile agrees very well with the MCAK current profile even for $k=4$. It must be noted that the system-size fluctuates in the EPFB model whereas it remains constant in the MCAK model supplemented by the rod-to-monomer mapping, Eq.(\ref{Eq:k-Curr}); yet, they seem to be equivalent as far as the steady-state density and current are concerned.\\

The EPFB model introduced here may be redefined in such a way that it generalizes the Dynamically Extending Exclusion Process (DEEP) of Sugden \emph{et al}., \cite{Sugden2007} in which the exiting monomer adds an extra site to the lattice. Let the position of a $k$-mer be the site occupied by its right-end. Adopting the same dynamical rules as the DEEP, a $k$-mer enters the lattice from the right, at a rate $\alpha$, if the last $k$ sites are empty; on entry, its position is $N$. Similarly, a $k$-mer at the site $k$ can exit from the left at a rate $\gamma$. In the bulk, a $k$-mer can jump to an empty nearest neighbor site to its left at a rate 1. In the corresponding EPFB model of representative monomers, the dynamical rules are the following:
\begin{eqnarray}
0 & \longrightarrow & 1 \quad \mbox{at site N, rate $\alpha$} \nonumber \\
1 & \longrightarrow & 0 \quad \mbox{in the bulk, rate 1} \nonumber \\
1 & \longrightarrow & 000...0 \quad \mbox{at site 1, at rate $\gamma$} \nonumber
\end{eqnarray}
The exit of a monmer in the EPFB model adds $(k-1)$ vacant sites to the lattice. Since the leftmost site is labelled as site 1, the size of the lattice $N$ grows indefinitely. Sugden \emph{et al}., \cite{Sugden2007} have considered the case when an exiting monomer adds one extra site to the lattice. They have shown that the maximal current in their model is $J = 3 - 2\sqrt{2} \equiv (\sqrt{2}+1)^{-2}$ which is what we would expect for hard-core dimers \cite{Lakatos2003}. Similarly, we may expect the maximal current in this generalized version of the DEEP model (may be named $k$-DEEP model) to be that for hard-core $k$-mers, namely $J = (\sqrt{k}+1)^{-2}$. An important difference between the $k$-DEEP model and the EPFB model discussed in this paper is that while the rod-to-monomer mapping given by Eqs.(\ref{Eq:k-Dens},\ref{Eq:k-Curr}) needs to be used in the EPFB model in order to obtain the steady state current and density of $k$-mers, such a mapping does not seem needed in the $k$-DEEP model, at least for non-interacting ($v=0$) particles. \\

Interestingly, the phenomenology of the EPFB model suggests that it could be considered as a simple model mimicking the unidirectional motion of a cell or a caterpillar that involves an alternating sequence of 'extensions' and 'contractions'. For example, instead of always labeling the left-end site entered by a monomer as the first site of the lattice, we may treat it as a variable. That is to say, whenever a monomer enters the system, the left-end site it occupies, say $j_l$, is updated to $j_l+(k-1)$. The right-end site, say $j_r$, from which a monomer exits the system is updated to $j_r+(k-1)$. So, the lattice 'contracts' when a monomer enters the system, and 'extends' when a monomer exits. Since these two events are not simultaneous in this model, the bulk of the system may be imagined as glued to the substrate while it extends or contracts. During the 'internal' process (hopping of monomers in the bulk), the lattice-size remains constant. Due to this alternating 'shrink-and-grow' process, the 'system' as a whole moves to the right; here, by 'system', we mean the lattice and the particles hopping on it. \\

\begin{figure}  
\begin{center}

 \includegraphics[width=0.7\textwidth]{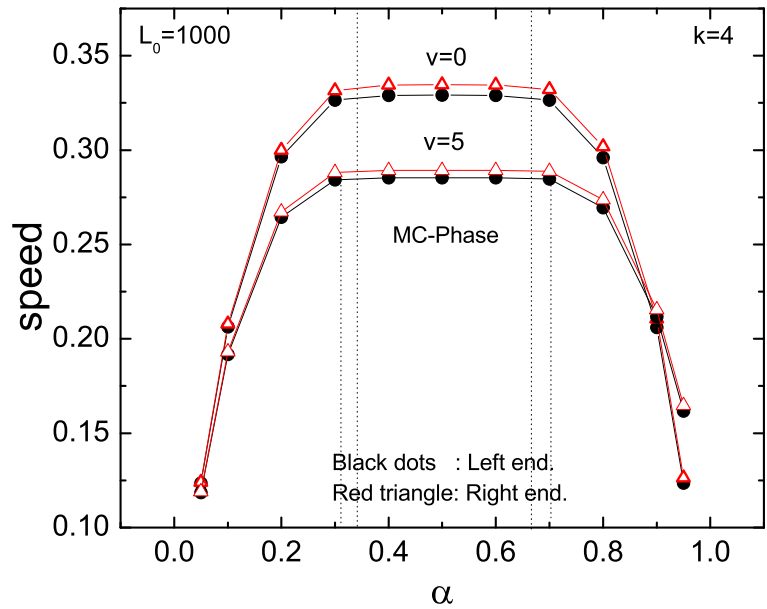}
  \caption{The speed with which the left and the right boundaries move for a sytem of non-interacting ($v=0$) as well as interacting ($v=5$) monomers (representing tetramers)  in the BFEP model. Initial size of the lattice $L_0=1000$; the data are an average of 50 runs, each consisting of $10^7$ moves. Black dots are for the left-end, while Red triangles are for the right-end. The small difference in speeds is due to the fact that counting of the entry and the exit events in the algorithm was done through the intial stages when the system was not in a steady state. } 
\label{Fig:Speed-BFEP}   

\end{center} 
\end{figure}

The net displacement of the left (right) boundary is simply the number of entry (exit) events multiplied by $(k-1)$; therefore, dividing it by the total Monte Carlo time in the EPFB algorithm gives an estimate of the speed of the corresponding boundary. In Fig.(\ref{Fig:Speed-BFEP}), we have presented the average speed of the boundaries for a system of non-interacting monomers (representing teramers). The system moves the fastest in the maximal current phase in the case of both $v=0, 5$. The small difference in speeds between the left-end and the right-end seen in this figure is due to the fact that counting of the entry and the exit events in the algorithm was done through the intial stages also when the system was not in a steady state. We have checked that they agree, within statistics, if the counting were done only after the system has reached the steady state. The system can also be made to move as a whole even in the case of monomers ($k=1$) if we define the size of the lattice as $L_0 - Nk$ in stead of $L_0 - N(k-1)$ in the EPFB model, and update the positions and numbering of the particles appropriately as well. It must be stressed here that the EPFB model only mimicks the unididirectional motion of a cell; it does not model the biophysical mechanism underlying this type of motion.   \\ 

In summary, we have demonstrated that the TASEP of an interacting system of $k$-mers on an open lattice is equivalent to the TASEP of interacting monomers on an open lattice with fluctuating boundaries. \\

$^3${present address : Departamento de F\'{\i}sica, Universidad de Extremadura, E-06071 Badajoz, Spain}

\bibliography{asep-bib.bib}
\bibliographystyle{plain} 


\end{document}